\documentclass[sigplan]{acmart}
\AtBeginDocument{%
  }

\usepackage{placeins}
\usepackage[english]{babel}
\usepackage{subcaption} 

\copyrightyear{2025}
\acmYear{2025}
\setcopyright{acmlicensed}\acmConference[BDCAT '25]{IEEE/ACM 12th International Conference on Big Data Computing, Applications and Technologies}{December 1--4, 2025}{Nantes, France}
\acmBooktitle{IEEE/ACM 12th International Conference on Big Data Computing, Applications and Technologies (BDCAT '25), December 1--4, 2025, Nantes, France}
\acmDOI{10.1145/3773276.3774296}
\acmISBN{979-8-4007-2286-8/2025/12}
\setcopyright{none}           
\begin{document}

\title{How to evaluate NoSQL Database Paradigms for Knowledge Graph Processing}

\author{Rosario Napoli}
\orcid{0009-0006-2760-9889}
\email{rnapoli@unime.it}
\affiliation{%
  \institution{University of Messina}
  \country{Italy}
  }
  
\author{Antonio Celesti}
\orcid{0000-0001-9003-6194}
\email{acelesti@unime.it}
\affiliation{%
  \institution{University of Messina}
  \country{Italy}
  }

\author{Massimo Villari}
\orcid{0000-0001-9457-0677}
\email{mvillari@unime.it}
\affiliation{%
  \institution{University of Messina}
  \country{Italy}
  }

\author{Maria Fazio}
\orcid{0000-0003-3574-1848}
\email{mfazio@unime.it}
\affiliation{%
  \institution{University of Messina}
  \country{Italy}
  }


\renewcommand{\shortauthors}{R. Napoli, A. Celesti, M. Villari, M. Fazio}

\renewcommand{\qedsymbol}{}

\begin{abstract}
Knowledge Graph (KG) processing faces critical infrastructure challenges in selecting optimal NoSQL database paradigms, as traditional performance evaluations rely on static benchmarks that fail to capture the complexity of real-world KG workloads. Although the big data field offers numerous comparative studies, in the KG context DBMS selection remains predominantly ad-hoc, leaving practitioners without systematic guidance for matching storage technologies to specific KG characteristics and query requirements. 
This paper presents a KG-specific benchmarking framework that employs connectivity density, scale, and introduces a graph-centric metric, namely Semantic Richness (SR), within a four-tier query methodology to reveal performance crossover points across Document-Oriented, Graph, and Multi-Model DBMSs. We conduct an empirical evaluation on the FAERS adverse event KG at three scales, comparing paradigms from simple filtering to deep traversal, and provide metric-driven, evidence-based guidelines for aligning NoSQL paradigm selection with graph size, connectivity, and semantic richness.
\end{abstract}

\ccsdesc[300]{Information systems~Data analytics}

\ccsdesc[500]{Information systems~Decision support systems}

\keywords{Graph Data Science, Knowledge Graphs, Knowledge Engineering, Graph Analytics, Performance Analysis.}


\maketitle
\section{Introduction}

The exponential growth in heterogeneous data generation has driven the widespread adoption of Knowledge Graphs (KGs) as  fundamental data structures for organizing and representing semantically enriched information \cite{b42}. KGs extend traditional graph structures by incorporating typed entities and relationships with rich metadata, enabling high-level representations of real-world dynamics. They are widely used in different domains, such as biomedical research or financial services, 
and their increasing relevance spans from enterprise data lakes \cite{b23} to semantic reasoning systems,  as ground truth sources for large language models (LLMs), where they help mitigate hallucination issues \cite{b21}.

The wide adoption of KG-based applications has given rise to Graph Data Science (GDS), a discipline focused on extracting insights from graph-structured data through advanced graph algorithms and KG-based Machine Learning (ML) techniques \cite{b29}. However, the efficient deployment of GDS solutions is affected by the selection of the most suitable Database Management System (DBMS) for data storage and querying. In fact, unlike traditional relational data, KGs present unique challenges, including heterogeneous node types, variable relationship densities, and complex traversal patterns, that demand storage systems optimized for the specific data features of the KG~\cite{b26}. This paper aims to address this problem by providing practical guidelines for selecting the DBMS that best fits the needs of a specific KG. 

The landscape of NoSQL DBMSs offers multiple paradigms, such as document-oriented, graph-native, and multi-model approaches, each one optimized for different data characteristics and query patterns \cite{b20}. In particular, document-oriented databases, like MongoDB\footnote{https://www.mongodb.com/}, are specialized in attribute-based filtering and nested data structures, while graph databases, such as Neo4j\footnote{https://neo4j.com/}, are purpose-built for relationship traversal and pattern matching. Multi-model systems, such as ArangoDB \footnote{https://arangodb.com/}, attempt to bridge these paradigms, offering flexibility at the cost of potential performance trade-offs \cite{b25}. However, identifying which DBMS paradigm performs better for specific KG characteristics and workload patterns remains a challenge, made even more complex by the continuously evolving landscape of big data management solutions. 
Inefficient storage decisions can lead to query response times that degrade exponentially with the KG size, scalability bottlenecks can limit analytical capabilities, and development overhead can consume resources from core GDS tasks \cite{b11}. Moreover, the heterogeneous nature of KGs, from small, semantically rich datasets to sparsely connected large-scale networks, demands a deep understanding of how different DBMS paradigms perform across this spectrum.

Currently, the selection of the DBMS for a KG is mainly made ad-hoc, relying on developer intuition or vendor marketing claims~
\cite{b30}. Studies on this topic in the literature carry out performance analyses of DBMSs based on synthetic datasets or single-domain applications. However, these approaches do not specifically address the requirements of KGs and leave practitioners without systematic guidance on how to select appropriate storage technologies based on the specific characteristics of their datasets and query requirements~\cite{b1}. 
To fill this gap, we present an analysis of NoSQL DBMS schemas' performance when handling KG workloads. Our results reveal significant performance variations across DBMS paradigms in the KG domain, with findings that can provide graph data scientists with concrete, empirically-grounded guidance for infrastructure decisions, potentially reducing development time and improving system performance across KG applications. In particular, this paper presents the following contributions:

\begin{enumerate}
    \item Systematic Multi-Dataset Evaluation: we conduct a reproducible KG-based comparative analysis of three key KG-suited NoSQL paradigms, that are document-oriented, graph-native, and multi-model, using real-world KGs with different features and properties.
    
    \item Progressive Query Complexity Analysis: we design and implement a four-tier query benchmark that progressively tests system capabilities from specific attribute filtering to complex multi-hop traversals, enabling identification of performance crossover points where optimal DBMS schema choice shifts.
    
    \item Practical Selection Guidelines: based on the achieved empirical results, we derive evidence-based recommendations considering three key KG dataset metrics:
    \begin{itemize}
        \item \textbf{Scale}: the volumetric complexity of the KG (\(|V| + |E|\)), which impacts storage and indexing overhead.
        \item \textbf{Connectivity}: the average relationship density (\(|E|/|V|\)), which affects traversal and join performance.
        \item \textbf{Semantic Richness}: the schema heterogeneity, which combines the number of distinct entity and relationship types with their distributional entropy.
    \end{itemize}
\end{enumerate}

The remainder of this paper is organized as follows. Section~\ref{sec:definitions} introduces the core definitions used throughout the paper. In Section~\ref{sec:background} we provide background on NoSQL DBMS for KG management while Section~\ref{sec:related} discusses related work on NoSQL database benchmarking in the KG context. 
In Section~\ref{sec:comparative} we present a comparative analysis of candidate DBMS schemas.
Section~\ref{sec:metrics} introduces the proposed complexity metrics for KGs, and Section~\ref{sec:methodology} describes the  methodology. 
The datasets, benchmark design, and execution setup are reported in Section~\ref{sec:experiments}, followed by the results and their discussion in Section~\ref{sec:results}. 
Finally, Section~\ref{sec:conclusions} concludes the paper and outlines directions for future research.

\section{Definitions} \label{sec:definitions}
This section introduces the formal definitions of core concepts used throughout the work.

\begin{definition}[Knowledge Graph]
A Knowledge Graph is defined as a tuple \( KG = (V, R, E, \ell_V) \), where:
\begin{itemize}
    \item \( V \) is the set of entities (nodes);
    \item \( R \) is the set of relationship types;
    \item \( E \subseteq V \times R \times V \) is the set of labelled, directed edges (triplets);
    \item \( \ell_V : V \rightarrow C \) is a labeling function that assigns exactly one class label from the set \( C \) to each node.
\end{itemize}
\end{definition}

\begin{definition}[KG atomic unit]
The atomic unit of a KG is a triple\cite{b32}, which consists of:
\begin{itemize}
    \item  labelled head node \( h \);
    \item a labelled tail node \( t \);
    \item a directed labelled edge \( r\) from \( h \) to \( t \).
\end{itemize}
\end{definition}

\begin{figure}[ht!]
    \centering
    \includegraphics[width=0.75\linewidth]{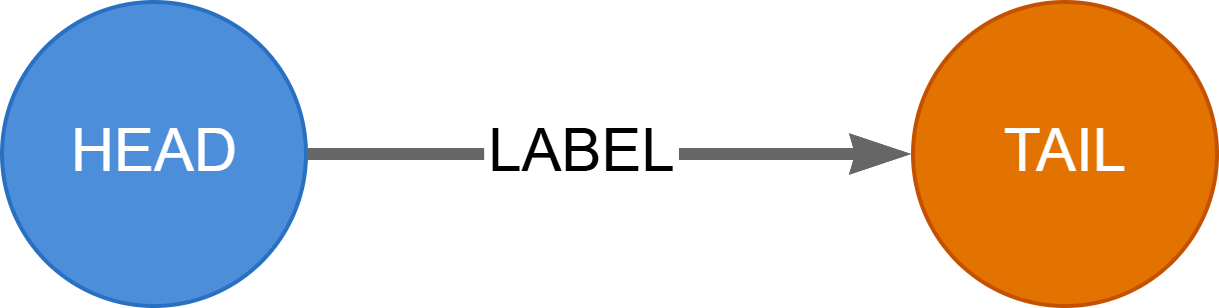}
    \caption{KG Atomic Unit.}
    \label{fig:placeholder}
\end{figure}

\begin{definition}[Cold-Start Condition]
Query execution without caches or prior optimizations: each query starts from empty memory, giving unbiased and reproducible performance measurements.
\end{definition}

\begin{definition}[Hot-Start Condition]
Query execution after warm-up: caches and optimizations from previous queries are used, reflecting steady-state performance with typically faster responses.
\end{definition}

\section{Background on NoSQL DBMS for KG Management}\label{sec:background}

The heterogeneous structure and semantics of KGs require highly adaptable approaches for storage and information retrieval. These techniques must consider not only the coexistence of multiple data types, but also scalable infrastructures that can efficiently handle dynamic graph growth and fluctuating query loads \cite{b1}. Such needs have encouraged the shift towards paradigms inspired by NoSQL databases, which were specifically designed to overcome the performance and scalability limitations of traditional relational database systems. 

Within this scenario, selecting a NoSQL DBMS in KG domain is not a simple performance-driven decision but it is also related to semantic expressiveness, since each paradigm provides different degrees of support for multi-hop traversal and heterogeneous schema representation. Several studies have analyzed how different DBMS architectures model and manage data, with particular focus on scenarios involving data integration from multiple sources \cite{b4}. Among NoSQL solutions, four main categories can be distinguished:
\begin{itemize}
    \item \textbf{Key-Value stores}, which manage information as simple pairs of keys and values;
    \item \textbf{Column-Oriented stores}, which store data by column, enabling selective access to only the attributes relevant to a query; 
    \item \textbf{Document-Oriented stores}, which manage semi-structured entities, such as JSONs; 
    \item \textbf{Graph-Oriented stores}, which natively model data as nodes and edges to explicitly represent relationships \cite{b18}.
\end{itemize}

Each category provides a different balance between schema flexibility, scalability, and support for complex queries. As a consequence, the choice of the underlying system plays an important role in KG management, as it directly affects the efficiency, scalability, and quality of knowledge extraction and reasoning in Graph Data Science (GDS) workflows \cite{b3}. Additionally, the rise of hybrid approaches that combine multiple NoSQL paradigms has introduced further complexity in the selection process \cite{b22}.

\begin{figure}[ht!]
\centering
\includegraphics[width=0.75\linewidth]{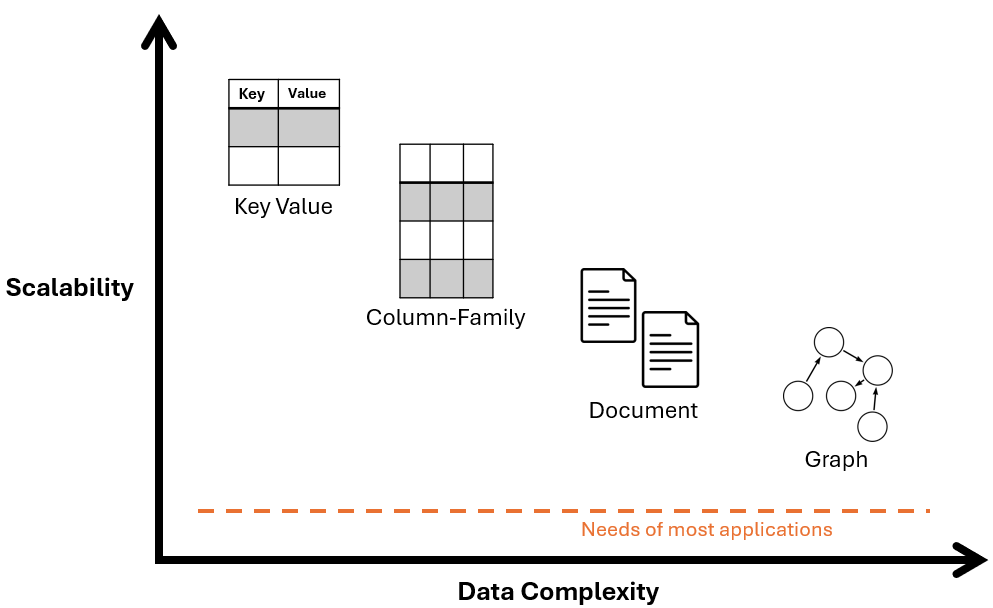}
\caption{Classification of NoSQL paradigms in terms of scalability and complexity.}
\label{fig:complexity_scalability}
\end{figure}

To better capture these differences, NoSQL DBMSs can be mapped along two key axes related to their internal schema model: scalability and complexity (see Fig. \ref{fig:complexity_scalability}). This mapping can help to understand how each paradigm fits KG modeling requirements, where large-scale data and semantically rich, interconnected structures often need to coexist.

Based on this classification, we now examine the main NoSQL paradigms in more detail, starting with Key-Value stores, which are particularly efficient for high-performance key-based lookups, especially in distributed environments. Nevertheless, their application in KG scenarios is limited, as they do not provide native support for relationships or semantic context. Representing a KG in such systems usually requires manually encoding graph semantics within keys or values, an approach that rapidly becomes inefficient. Furthermore, executing multi-hop queries or pattern matching is impractical without reconstructing the graph externally, reducing their usability for KG-related tasks \cite{b24}.

Column-family databases, such as Cassandra\footnote{https://cassandra.apache.org/index.html}, offer the possibility to store wide rows with variable columns, which can be used to encode adjacency matrices to represent graphs. Despite this advantage, they still lack native mechanisms for efficient edge traversal and semantic inference \cite{b19}. As a result, column-family models share similar limitations with key-value stores when it comes to expressive relationship querying.

In contrast, graph databases and document stores have emerged as particularly suitable for KG management due to their ability to handle heterogeneous information and optimize specific query patterns \cite{b8}. In particular, document-oriented databases allow nested entities and cross-document references, enabling the representation of local graph fragments inside a single document or collection. 
Graph databases, on the other hand, are specifically engineered to store and query networks of interconnected elements efficiently but are poorly suited to handle vast amount of data.

Finally, the emergence of multimodel data management solutions has introduced an additional layer of complexity, as they aim to combine the benefits of multiple NoSQL paradigms under a single framework, 
further complicating the decision-making process for KG management.

In summary, NoSQL databases represent an important foundation for KGs representation, as their ability to deal with heterogeneous and evolving datasets, together with their scalability in distributed settings, supports efficient reasoning over complex structures. Nonetheless, the impact of these benefits relies heavily on choosing the most suitable paradigm, as each comes with specific trade-offs that influence KG performance, scalability, and overall usability.

\section{Related Work}\label{sec:related}

Performance evaluation of NoSQL database systems has emerged as a significant research area, driven by the need to understand how different storage paradigms handle multiple workloads and data characteristics. Although there are extensive benchmarking studies for the traditional landscape of NoSQL systems, KG applications in the NoSQL domain must analyze specific trade-offs related to heterogeneous architectures and specialized optimization strategies \cite{b27}. 

However, most existing performance studies focus only on general database workloads rather than the specific requirements of KGs management \cite{b6}. In particular, KGs present different challenges, including heterogeneous node types, variable relationship densities, complex traversal patterns, and semantic reasoning requirements that require specialized evaluation methodologies \cite{b33}.

Furthermore, the KG community has not yet undergone a full transition from classical approaches, such as Resource Description Framework (RDF), a W3C standard that represents knowledge as subject–predicate–object triples, toward more modern NoSQL-based methodologies \cite{b44}. This partial adoption creates a fragmented landscape in which practitioners lack clear guidelines for selecting the most appropriate data model and database technology. As a result, graph data scientists usually rely on their own prior knowledge or expertise when making architectural choices, leading to inconsistent practices across projects.

This gap poses significant challenges for scalable KG deployment, as suboptimal DBMS architecture decisions can lead to performance issues that increase with dataset growth and query complexity. This forces practitioners to invest substantial effort in database system evaluation and optimization rather than focusing on graph algorithm development and analytical insights, a resource allocation inefficiency that limits innovation in graph data science workflows.

\subsection{NoSQL Performance Benchmarking}

Traditional NoSQL benchmarking approaches primarily target generic data models, focusing on CRUD operations, throughput metrics, and scalability under uniform workloads \cite{b32}. Early comparative studies provided performance baselines for document stores, key-value systems, and graph databases using domain-agnostic large-scale datasets. However, these methodologies are not designed for KG scenarios, as they ignore the semantic richness and relationship complexity that characterize them \cite{b34}.

Recent benchmarking efforts have also moved towards incorporating more realistic workloads, including multi-model scenarios and heterogeneous query patterns. In particular they shown significant performance variations across NoSQL paradigms when executing complex relational queries, with comparative analyses of different paradigms that reveal substantial differences in traversal efficiency and pattern matching capabilities \cite{b8}.
Nevertheless, these studies are still constrained by their reliance on generic Big Data datasets. As a result, they fail to capture the heterogeneity, semantic richness, and relationship complexity that define real-world KG scenarios, limiting the applicability of their findings in KG-oriented use cases.

\subsection{Knowledge Graph Storage Systems}

The intersection of KG management and storage system performance represents a relatively new research area. Early work focused primarily on RDF triple stores and their optimization for SPARQL queries, establishing performance patterns for semantic web applications \cite{b1}. In particular, traditional approaches for KG representation have largely focused on RDF triple stores and relational schema-based models, emphasizing SPARQL query performance and reasoning capabilities that have shaped much of the early research on KG storage and querying \cite{b17}. Although NoSQL systems offer alternative data modeling strategies, there is still limited work examining how these systems handle the structural and semantic complexity typical of KGs.

In particular, a critical research gap persists regarding how the intrinsic properties of datasets, such as semantics, connectivity density, and scalability requirements, affect the suitability of NoSQL paradigms for KG representation. Existing studies tend to either replicate traditional triple-store settings, leaving the interaction between data characteristics and NoSQL storage strategies underexplored \cite{b36}.
However, systematic comparative studies across multiple real-world datasets with different characteristics remain limited, leaving practitioners without evidence-based guidance for storage system selection.

\section{DBMS Schemas for KG representation}\label{sec:comparative}

To better understand how current database technologies support the specific needs of KG management, we will focus on data models that are commonly considered more expressive and evolved for representing KGs. For these reasons, we exclude key-value and column-oriented databases. Although these models can, to some extent, store graph-like structures (e.g., by encoding adjacency or entity attributes), their lack of native mechanisms for edge traversal, semantic inference, and expressive relationship querying makes them inherently less suitable for the advanced reasoning and high-level data interlinking.
In fact, most of the existing literature on KG data management concentrates on solutions that are explicitly designed to support heterogeneous knowledge structures and graph-oriented reasoning \cite{b45}. Graph databases, document-oriented systems, and multi-model approaches are consistently identified as the most promising paradigms, since they offer mechanisms that can balance flexibility, semantic richness, and scalability.

This decision allows us to investigate how the most suitable NoSQL paradigms address the core challenges of managing not only highly interconnected but also heterogeneous data, where nodes may belong to different classes and exhibit numerous attributes, offering a clearer understanding of their respective strengths and trade-offs in practical KG scenarios.
In particular, we examine three representative systems: the graph database Neo4j, the document-based MongoDB, and the hybrid graph-document ArangoDB. Each of these represents a distinct NoSQL paradigm frequently employed in knowledge-driven contexts. The following subsections explore their architectural foundations and design choices in relation to KG-specific requirements.

\subsection{Neo4j: Native Graph Modeling for Knowledge Graphs}
Neo4j is widely recognized as one of the most established technologies for graph data management and it is frequently adopted in KG research and applications \cite{b9}. Based on the graph model, it enables the direct representation of entities and their semantic relationships as nodes and edges, each with key-value properties. This native graph structure is particularly well-suited for capturing the expressiveness and connectivity inherent in KGs.

One of the defining features of Neo4j is its declarative query language, Cypher, which is specifically designed for handling complex graph patterns. It supports multi-hop traversal and subgraph matching, which are fundamental operations in KG exploration and reasoning \cite{b10}. Unlike general-purpose NoSQL systems, Neo4j is built to handle graph-centric workloads, making it highly suitable for domains that require modeling rich ontologies, inferring implicit knowledge, and navigating deeply nested relationships. However, this  architecture comes with scalability limitations, as graph DBMSs exhibit significant performance degradation when processing large-scale datasets, with query execution times increasing exponentially as graph size and complexity grow (Fig. \ref{fig:complexity_scalability}) \cite{b11}.

\subsection{MongoDB: Document-Oriented Flexibility for Semi-Structured Knowledge}
MongoDB is the most widely used document-oriented NoSQL database and provides a flexible schema design based on semi-structured representations of knowledge \cite{b14}. Although it does not natively support graph operations, its usage of a JSON-like format allows for the nesting of documents and hierarchical structuring, which can be adapted to capture portions of KG-like information in the database.

Specifically, it organizes data into collections of documents, each of which may embed multiple layers of related information. This approach is beneficial in applications where knowledge is locally structured and data relationships can be encoded within the document hierarchy \cite{b15}. 
In particular, the document model can be seen as a superset of other data models including key-value, relational, objects and graph.
Specifically, it provides a graphLookup stage within its aggregation pipeline, which enables recursive traversal of graph-like structures such as trees and hierarchical data. This feature supports the exploration of connected documents by performing repeated lookups in a specified relationship until a termination condition is met. However, graphLookup does not offer the full range of graph algorithms or the optimized traversal performance typically available in native graph databases.

In the end, this document-centric architecture presents inherent limitations for graph-intensive operations, MongoDB's performance for complex graph traversals degrades significantly compared to native graph databases, particularly when dealing with highly interconnected knowledge graph topologies that require extensive cross-document relationships and multi-hop queries (Fig. \ref{fig:complexity_scalability}).

\subsection{ArangoDB: Unified Access through a Multi-Model Design}
ArangoDB introduces a multi-model architecture that natively supports two different paradigms: document and graph within a single integrated engine. 
This design offers a unified interface for applications that require different types of data access without resorting to separate database systems or polyglot persistence.

In the context of KGs, ArangoDB provides native support for graph structures while allowing document-oriented storage. Its flexibility makes it suitable for modeling hybrid scenarios in which some knowledge is highly relational, whereas other information is better represented as attributes or nested entities \cite{b13}. The internal data representation is JSON-based, and data are grouped into collections that distinguish between document and edge types. 

It uses its own query language, AQL (Arango Query Language), which is inspired by SQL and supports operations across multiple models. Although AQL allows graph traversal, it is generally more verbose and less semantically focused than Cypher, making the user experience less intuitive in graph-heavy contexts.

Our comparative analysis highlights that none of the examined DBMS solutions can be considered ``perfect'' in absolute terms. These heterogeneous strengths and weaknesses confirm that the choice of the most suitable DBMS strictly depends on the structural and semantic characteristics of the KG, 
underscoring the need for an adaptive and systematic approach to support the selection process in knowledge graph management.

\section{KG Complexity Metrics for NoSQL DBMS Evaluation}\label{sec:metrics}

Our review of related work in the literature 
shows that existing benchmarks do not take into account critical KG‐specific factors. Traditional evaluations of NoSQL DBMSs primarily emphasize aggregate throughput, latency, and scalability metrics, which are valuable indicators in generic database contexts but are insufficient to capture the complexity of KGs. In particular, semantic relationships can be very different, producing unbalanced distributions where some regions of the graph are dense and semantically rich, while others are sparse and attribute-centric. 
This heterogeneity has a significant impact on DBMS performance. For instance, queries in a graph-native store may become computationally expensive if applied to large attribute payloads or long-tail distributions of node types, whereas a document store may outperform in attribute-heavy contexts but struggle with deep traversals or relationship-intensive queries. These considerations highlight the necessity of a systematic methodology that links measurable KG properties with observed system performance.

To fill this gap, we aim to evaluate three orthogonal and quantifiable dimensions of KGs: Scale, Connectivity, and Semantic Richness, each one denoting a specific aspect of KG complexity. These dimensions allow us to design a set of experiments for systematically exploring how different NoSQL paradigms respond to changes in KG topologies, schema distributions, and semantic configurations. By mapping performance results into this three-dimensional space, we can identify the crossover points at which one paradigm begins to outperform another. This provides practitioners with evidence-based guidelines for selecting a DBMS. 

Given a Knowledge Graph \(KG = \langle V, R, E, \lambda_V \rangle\) as previously defined, we introduce the following dimensions:

\begin{definition}[Scale]
Given \(KG\), the Scale dimension \(S(KG)\) measures its volumetric complexity:
\[
S(KG) \;=\; |V| + |E|.
\]
Here \(|V|\) and \(|E|\) denote the numbers of entities and edges, respectively.  
Scale can be used for storage requirements, indexing overhead, and memory consumption. 
\end{definition}

\begin{definition}[Connectivity Density]
Given \(KG\), the Connectivity dimension \(CD(KG)\) quantifies average relationship density:
\[
CD(KG) \;=\; \frac{|E|}{|V|}.
\]
Higher \(CD(KG)\) indicates more interconnected graphs, increasing traversal complexity and join operations.
\end{definition}

\begin{definition}[Semantic Richness]
  
Given \(KG = \langle V, R, E, \lambda_V \rangle\), the Semantic Richness \(SR(KG)\) measures how varied and expressive a graph’s schema is by combining two aspects: the number of distinct elements and the homogeneity of their distribution.

First, let:
\begin{itemize}
    \item $|C| = \text{number of entity types}$
    \item $|R| = \text{number of relationship types}.$
\end{itemize}

We capture type diversity via:
\[
D_{\text{types}} = \log|C| + \log|R|.
\]
A larger \(D_{\text{types}}\) means more different kinds of nodes and edges.

Next, we measure distribution uniformity using Shannon entropy, which quantifies the degree of uncertainty or randomness in the data distribution \cite{b46}. Formally:
\[
p(c) = \frac{\bigl|\{v\in V: \lambda_V(v)=c\}\bigr|}{|V|},
\quad
p(r) = \frac{\bigl|\{(u,r,v)\in E\}\bigr|}{|E|}.
\]
With $(u,r,v)$ the atomic units of KGs.
Then:
\[
H(C) = -\sum_{c\in C} p(c)\,\log p(c),
\quad
H(R) = -\sum_{r\in R} p(r)\,\log p(r).
\]
Higher entropy means types or relationships are more evenly represented, increasing query complexity.

Finally, we combine these components:
\[
SR(KG) = D_{\text{types}} + H(C) + H(R)
\]

In particular, \(SR(KG)\) grows when a graph has:
\begin{itemize}
  \item many different node and edge types (\(D_{\text{types}}\) large);
  \item and/or a balanced distribution of those types (high entropy).
\end{itemize}

\end{definition}

These dimensions collectively enable:
\begin{enumerate}
  \item \textbf{Dataset Selection:} Choosing evaluation graphs that span diverse regions of the \((S,CD,SR)\) space ensures coverage of real-world KG scenarios.
  \item \textbf{Performance Interpretation:} Mapping observed DBMS performance to specific \((S,CD,SR)\) combinations reveals crossover points where one paradigm outperforms another.
  \item \textbf{Generalization:} Deriving selection guidelines that apply beyond individual datasets by relating performance trends to underlying KG characteristics.
\end{enumerate}

By grounding our experimental methodology, we transform ad-hoc DBMS selection into a systematic, reproducible process that connects quantifiable KG properties to empirically observed performance patterns.  

\section{EVALUATION METHODOLOGY}\label{sec:methodology}

We propose a reproducible workflow to evaluate NoSQL DBMS paradigms on KG workloads across multiple scales. The methodology replicates the same process on datasets of progressively increasing dimensions, enabling systematic exploration of the scalability-complexity performance space (Fig. \ref{fig:complexity_scalability}). This approach allows us to observe how the three database paradigms behave under different conditions and identify performance inflection points as data volume grows exponentially.
The workflow consists of five sequential steps: Data Preparation, System Ingestion, Metric Computation, Query Execution, and Result Aggregation, executed identically across all datasets.
In the following subsections, we present a detailed analysis of the evaluation methodology, describing each component of our  approach.

\subsection{Data Preparation}
Input datasets are provided in graph format containing entity and relationship structures. We develop automated connectors capable of transforming and converting the original graph representation into the required target formats. Specifically, from the native graph structure, we generate JSON representations that are well-suited for document-oriented and multi-model systems.

\subsection{Dataset Scaling Strategy}
To evaluate system scalability under progressively larger workloads, we adopt an iterative duplication approach. 
Starting from the original dataset of size \(X\), we duplicate and merge the entire dataset with itself, effectively doubling its size at each step. 
After \(n\) duplications, the dataset size follows the exponential growth function:

\[
\text{Dataset Size After } n \text{ Duplications} = X \times 2^{n}
\]

\subsection{System Ingestion}
Each dataset undergoes a format-specific transformation before ingestion into the three paradigms. Graph-native systems receive direct ingestion of the original graph format, while Document-oriented and multi-model systems process data through a two-phase ingestion approach using automated connectors. The transformation process converts the original graph structure into two separate JSON representations: one with individual node classes and one containing RDF triples that encode subject-predicate-object structures, well-suited for document and multi-model paradigms.

\subsection{Metric Computation}
Immediately after ingestion, we compute the three KG complexity metrics on the imported data:
\begin{itemize}
  \item $S(KG) = |V| + |E|$;
  \item $CD(KG) = \frac{|E|}{|V|}$;
  \item $SR(KG) = D_{\text{types}} + H(C) + H(R)$.
\end{itemize}

\subsection{Query Execution and Result Aggregation}
We execute a suite of four query tiers to stress different system capabilities:
\begin{itemize}
  \item \textbf{Tier 1:} Complex single‐attribute filters.
  \item \textbf{Tier 2:} Compound multi‐attribute selections.
  \item \textbf{Tier 3:} Limited (1–2 hop) pattern matching.
  \item \textbf{Tier 4:} Deep (3+ hop) multi‐hop traversals.
\end{itemize}
Each query is run in two phases:
\begin{enumerate}
  \item \textbf{Cold‐start:} Clear caches before each of 31 runs.
  \item \textbf{Hot‐start:} Maintain caches across 31 runs.
\end{enumerate}
For each run, we measure \textbf{execution time} (milliseconds) and in the end, we aggregated the results of individual queries into charts that depict the performance of different DBMSs executing the same query on datasets of increasing size, plotting the mean values with \(95\%\) confidence intervals.

\section{EXPERIMENTAL RESULTS}\label{sec:experiments}

This section details the experimental workflow following the evaluation methodology. It covers the database systems tested, data preparation and scaling, ingestion procedures, computed metrics, query types, and the measured performance for each query. All experiments were conducted on a single workstation equipped with an Intel Core Ultra 5 processor, 16 GB of RAM, and an Intel Arc series GPU running under Windows 11 Pro.

\subsection{Dataset Preparation}
\begin{figure}[ht!]
    \centering
    \includegraphics[width=0.40\textwidth]{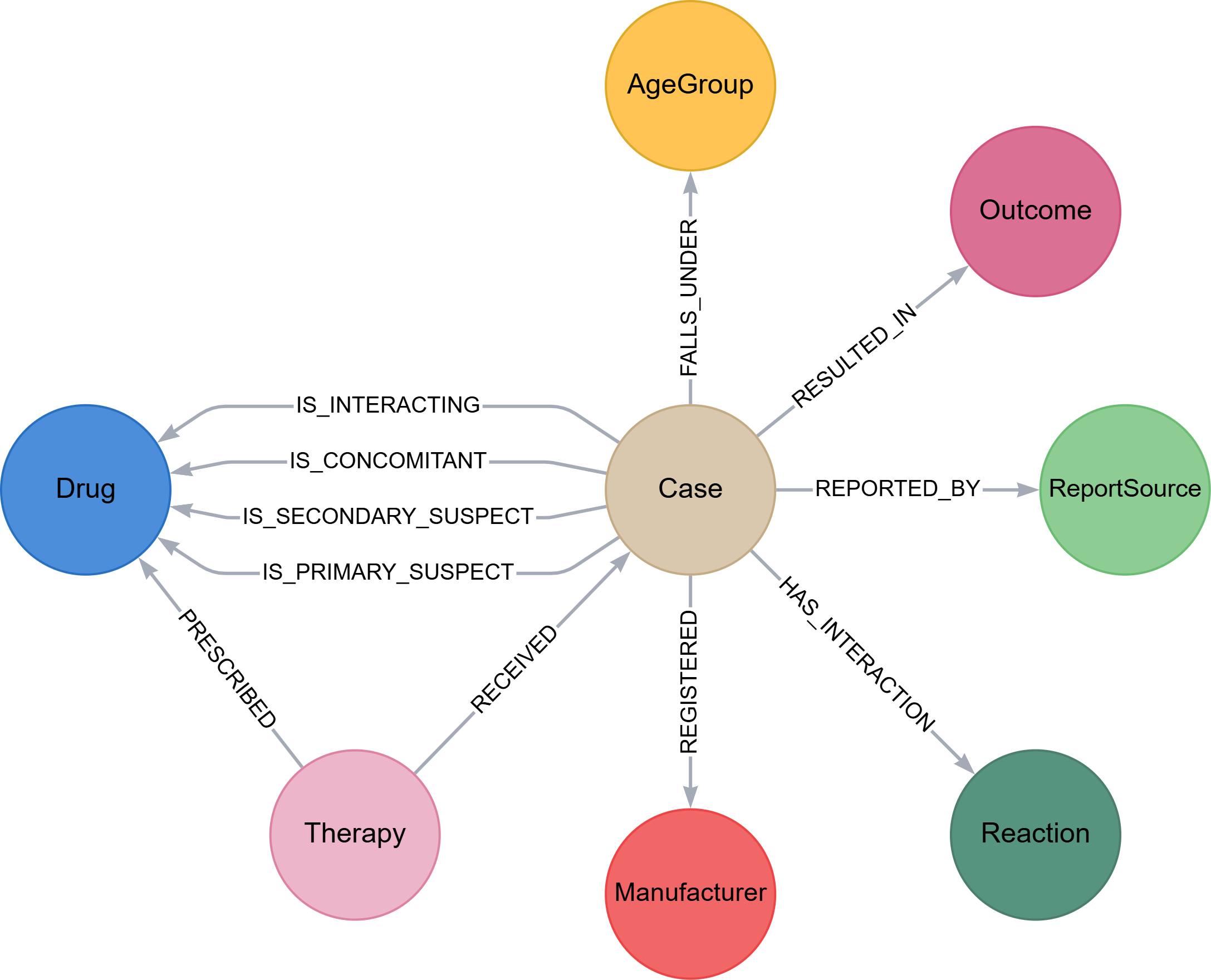}
    \caption{Schema of the FAERS Knowledge Graph.}
    \label{fig:faers_schema}
\end{figure}

Experiments were performed on a real-world dataset derived from the United States Food and Drug Administration (FDA) Adverse Event Reporting System (FAERS)\footnote{https://open.fda.gov/data/faers/}. FAERS is a database developed by the FDA to support post-marketing safety surveillance of approved drugs. In particular, it collects reports of adverse events and medication errors submitted by healthcare professionals, consumers and manufacturers. The dataset was formalized as a KG\footnote{https://github.com/neo4j-graph-examples/healthcare-analytics} in Neo4j with classes: Case, Drug, Reaction, Outcome, ReportSource, Therapy and Age Group (Fig. \ref{fig:faers_schema}), containing approximately 11,000 nodes and 14,500 edges across eight main entities. 

Data preparation began with exporting the entire graph into intermediate CSV files, one file for each node label and one for each relationship type. Each CSV record contains the element’s unique Neo4j identifier and a JSON‐serialized representation of its properties. The serialization routine handles nested dictionaries and lists. From these CSVs, two sets of JSON documents were generated:
\begin{itemize}
    \item For MongoDB, each node document uses the Neo4j ID and includes a properties field with the deserialized JSON object; relationship documents include: id, from,  to and "properties";
    \item For ArangoDB, a global lookup table maps Neo4j IDs to their labels, enabling construction of key, from and to fields in the format \verb|<collection>/<_key>|, with all original information preserved under "properties" field. 
\end{itemize}

\subsection{Dataset Scaling Strategy}
Starting from the original FAERS, we duplicate and merge the entire graph with itself at each step, yielding exponential growth. In particular, we selected three representative scales:
\begin{itemize}
  \item \textbf{1×}: the original dataset, baseline size.
  \item \textbf{3 duplications (8×)}: medium‐scale scenario stressing both ingestion and query performance.
  \item \textbf{7 duplications (128×)}: large‐scale scenario to push system limits and observe scalability trends.
\end{itemize}

\subsection{System Ingestion}

The ingestion layer orchestrates the transfer of prepared JSON representations into the target datastores using their native drivers. For MongoDB, a Python client establishes a connection to the specified database and iterates over each collection of node and relationship documents.

In the ArangoDB pipeline, the ingestion script leverages the ArangoDB Python driver to connect to both document and edge collections. Node documents are loaded into their respective vertex collections, and edge documents into the designated edge collections by \texttt{\_from} and \texttt{\_to} fields.

\subsection{Metric Computation}
Immediately after ingestion, we compute the knowledge graph complexity metrics on each scale configuration. Table~\ref{tab:srkg_metrics} reports comprehensive semantic richness metrics and connectivity measures computed across all scale configurations, providing detailed insights into the structural and semantic properties of the KG at different granularity levels.

\begin{table}[ht!]
    \centering
    \caption{Semantic richness metrics and connectivity density (CD) for FAERS KG at different scales.}
    \label{tab:srkg_metrics}
    \resizebox{\columnwidth}{!}{%
    \begin{tabular}{lrrrrrrr}
    \hline
    Scale & Nodes & Relationships & Dtypes & HC & HR & SR & CD \\
    \hline
    1×      & 14,000    & 11,000     & 4.48  & 1.39 & 2.04 & 7.91 & 0.79 \\
    8×      & 112,000   & 88,000     & 4.48  & 1.39 & 2.04 & 7.91 & 0.79 \\
    128×    & 1,792,000 & 1,408,000  & 4.48  & 1.39 & 2.04 & 7.91 & 0.79 \\
    \hline
    \end{tabular}%
    }
\end{table}

\subsection{Query Execution and Results Aggregation}

A Python script drives the execution of four query tiers in MongoDB, ArangoDB and Neo4j over the 31 cold and warm runs. The aggregation step reads the CSV files produced by the query execution phase, each containing per‐run timings and metadata for cold‐start and hot‐start executions. A Python script loads these CSVs, computes the mean execution time and the 95\% confidence interval across the 31 runs, merges it and plots them. In the following, we analyze the individual queries along with their aggregated results.

\begin{figure}[ht!]
    \centering
    \includegraphics[width=0.9\linewidth]{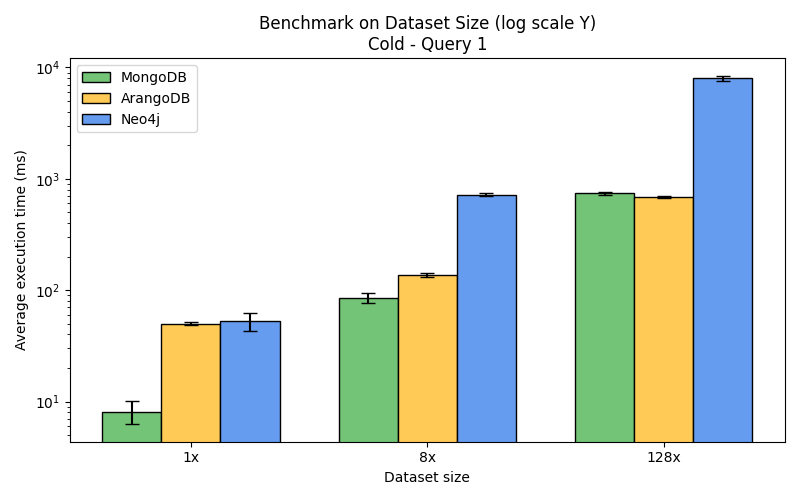}
    \caption{Cold‐start performance for Query 1 across systems.}
    \label{fig:query1_cold}
\end{figure}

\begin{figure}[ht!]
    \centering
    \includegraphics[width=0.9\linewidth]{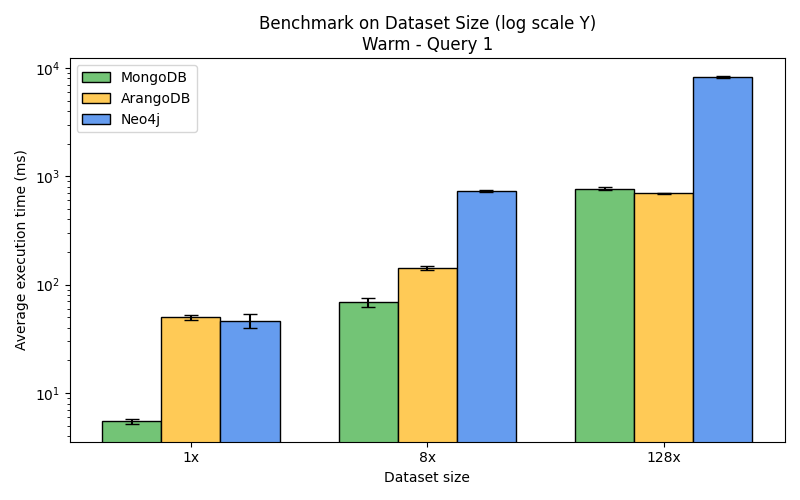}
    \caption{Warm‐start performance for Query 1 across systems.}
    \label{fig:query1_warm}
\end{figure}

\subsubsection{Tier 1: Complex Attribute-based Filtering (Figs. \ref{fig:query1_cold} and \ref{fig:query1_warm})}
In the first tier, we focus on narrowing down case reports by combining age, occupation, and gender criteria. The system scans every case record, selects those where the patient age falls between 60 and 90, the report originates from hospital staff, and the gender is marked as female or male. 


\begin{figure}[ht!]
    \centering
    \includegraphics[width=0.9\linewidth]{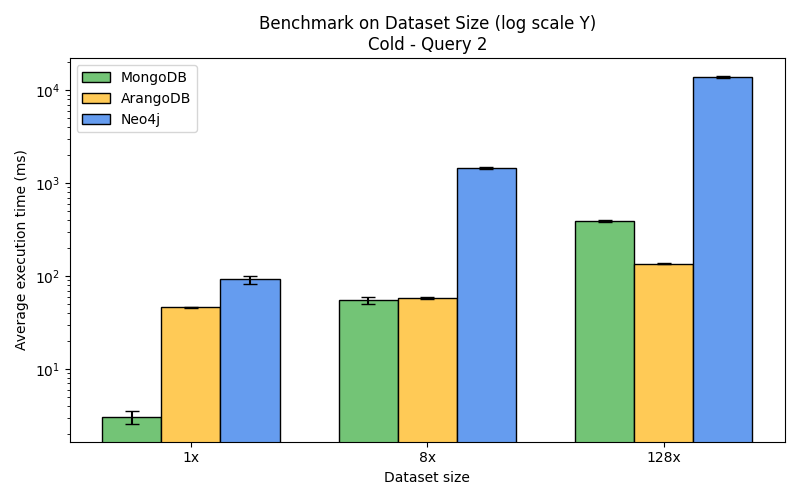}
    \caption{Cold‐start performance for Query 2 across systems.}
    \label{fig:query2_cold}
\end{figure}

\begin{figure}[ht!]
    \centering
    \includegraphics[width=0.9\linewidth]{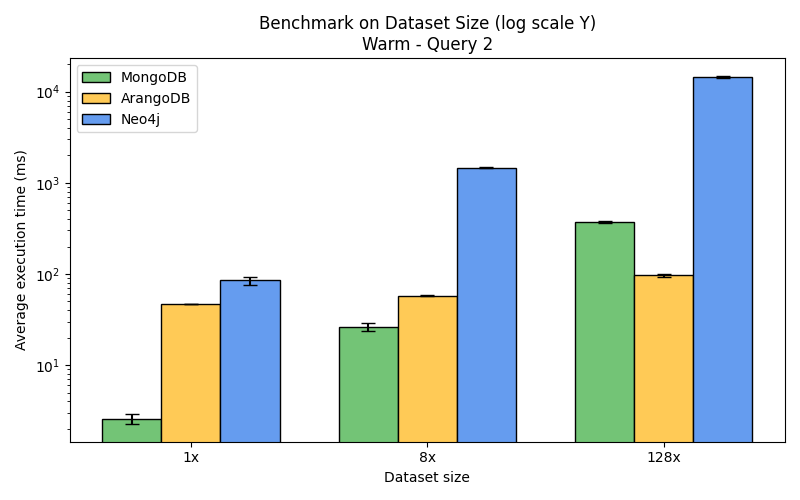}
    \caption{Warm‐start performance for Query 2 across systems.}
    \label{fig:query2_warm}
\end{figure}

\subsubsection{Tier 2: Filtering and 1-Hop Join (Figs. \ref{fig:query2_cold} and \ref{fig:query2_warm})}
In this scenario, the system first identifies all case reports submitted by hospital staff. For each of these cases, it follows all the relationship that link cases to their corresponding age group. Only those age groups labeled “Child” or “Adult” are retained. 


\begin{figure}[ht!]
    \centering
    \includegraphics[width=0.9\linewidth]{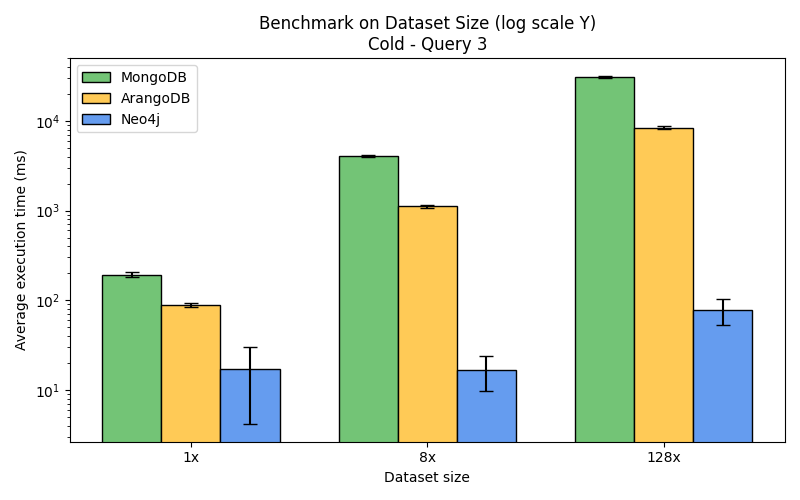}
    \caption{Cold‐start performance for Query 3 across systems.}
    \label{fig:query3_cold}
\end{figure}

\begin{figure}[ht!]
    \centering
    \includegraphics[width=0.9\linewidth]{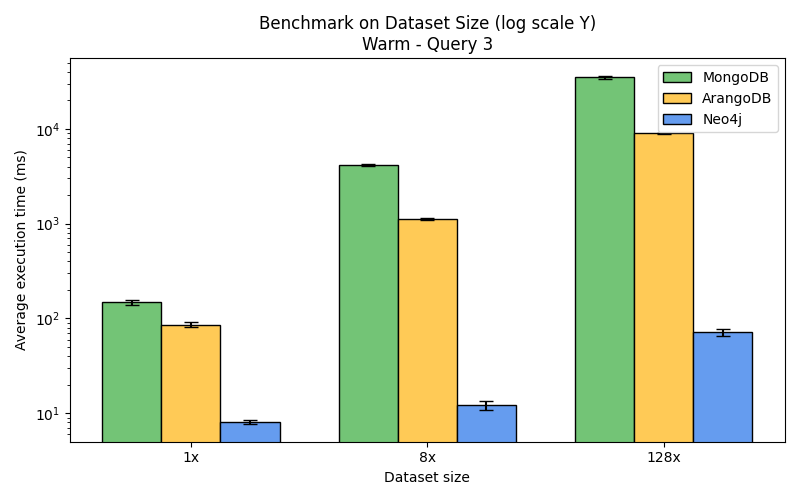}
    \caption{Warm‐start performance for Query 3 across systems.}
    \label{fig:query3_warm}
\end{figure}

\subsubsection{Tier 3: Multi‐Join Case to Manufacturer, AgeGroup, Drugs (Figs. \ref{fig:query3_cold} and \ref{fig:query3_warm})}

Here, the query starts by finding every case that has at least one prescribed drug. For each of those cases, it then follows two additional links: one to the manufacturer of the drug and another to the patient’s age group. The system assembles, for each case, the drug, its producer, and the age bracket of the patient. 


\begin{figure}[ht!]
    \centering
    \includegraphics[width=0.9\linewidth]{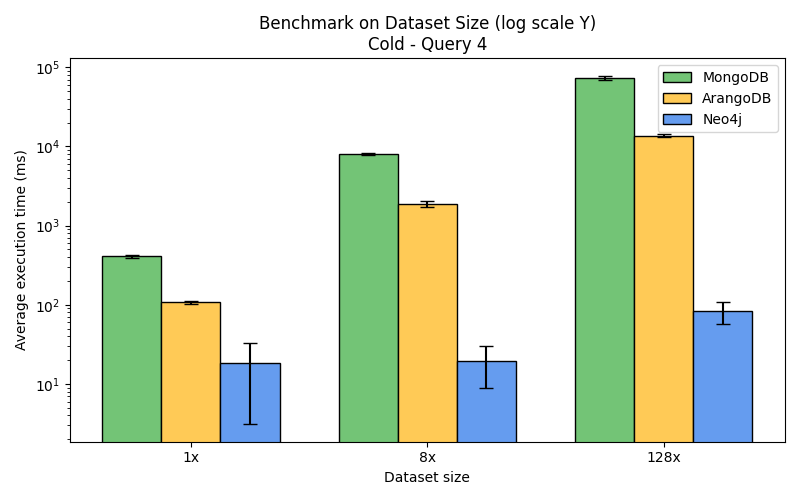}
    \caption{Cold‐start performance for Query 4 across systems.}
    \label{fig:query4_cold}
\end{figure}

\begin{figure}[ht!]
    \centering
    \includegraphics[width=0.9\linewidth]{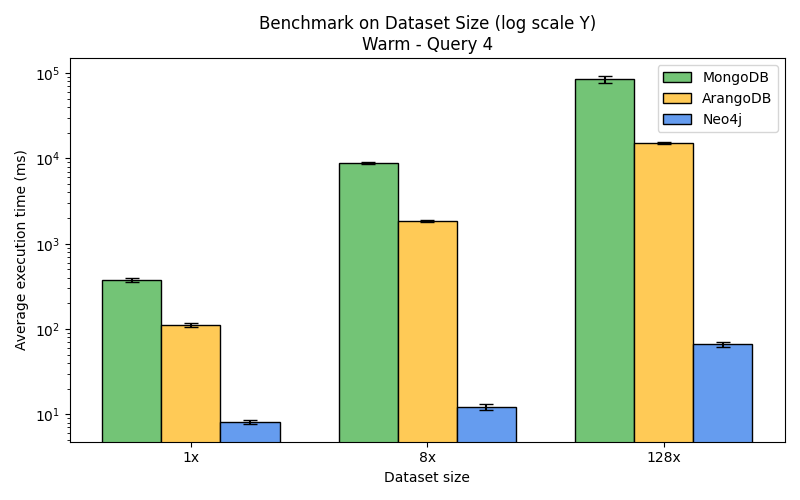}
    \caption{Warm‐start performance for Query 4 across systems.}
    \label{fig:query4_warm}
\end{figure}

\subsubsection{Tier 4: Deep Context Case and All Connected Entities (Figs. \ref{fig:query4_cold} and \ref{fig:query4_warm})}

The final query explores the full neighborhood around each case record. Beginning with each case node, the system retrieves every directly connected entity: manufacturer, age group, outcome, reporting source, drug, reaction, and therapy.


\section{Results Discussion} \label{sec:results}

The experimental results provide an integrated view of how NoSQL paradigms perform across a spectrum of queries while accounting for the semantic and structural properties of the FAERS KG. 

For Tier 1 attribute filters, MongoDB’s document‐oriented engine consistently yields the lowest response times (Fig.~\ref{fig:query1_cold}, Fig.~\ref{fig:query1_warm}). In this setting, SR$_{\mathrm{KG}}=7.91$ and C$_{\mathrm{KG}}=0.79$ imply moderate schema heterogeneity but limited connectivity, enabling efficient single‐collection scans and index seeks. As dataset scale increases from 1× to 128×, the linear growth in node and edge counts does not alter indexing performance, confirming document stores as the preferred choice for low‐hop, high‐throughput lookups. Moreover, when certain subgraphs contain rich entities but remain sparsely connected, such as outlier Drug or Outcome nodes with extensive metadata document‐oriented systems allow data scientists to perform deep attribute analyses without incurring graph‐traversal overhead. However, at the final scale step (128×), a very slight advantage begins to emerge in favor of ArangoDB, suggesting that its hybrid storage model may start to better balance index resolution and lightweight edge navigation under extreme data volumes.

Transitioning to Tier 2, which introduces hop joins, we confirm the previous trend and observe the performance shift favoring multi‐model databases as dataset size increases (Fig.~\ref{fig:query2_cold}, Fig. \ref{fig:query2_warm}). Here, the constant D$_{\mathrm{types}}=4.48$, $H(C)=1.39$, and $H(R)=2.04$ yield a balanced SR$_{\mathrm{KG}}$ that challenges pure document engines but remains below deep‐traversal thresholds. The Multi‐model systems alternate document retrieval with native graph navigation, absorbing moderate C$_{\mathrm{KG}}$ (0.79) without incurring full graph indexing overhead. This hybrid capability is particularly advantageous in KGs that exhibit heterogeneous connectivity, with regions of sparse but information‐rich nodes alongside densely connected clusters, allowing efficient localized traversals while preserving attribute‐centric performance across the graph.

In Tier 3 and 4 deep traversals, graph‐native platforms exhibit superior warm‐run performance (Fig.~\ref{fig:query3_cold} - Fig.~\ref{fig:query4_warm}) as dataset scale grows. The high SR$_{\mathrm{KG}}$ and C$_{\mathrm{KG}}$ amplify random I/O demands and multi‐hop path enumeration, which Neo4j addresses through adjacency‐optimized storage and native traversal planning. Cold‐start measurements (Fig.~\ref{fig:query4_cold}) reflect initial cache warming in the confidence intervals, but the warm‐run curves demonstrate that, for richly connected graphs with broad type diversity, such as comprehensive traversals, graph‐native engines deliver the most efficient execution of deep semantic queries.

The crossover points observed in the warm‐run plots of Figs.~\ref{fig:query1_warm}, \ref{fig:query2_warm}, and \ref{fig:query4_warm} delineate empirical boundaries in the SR$_{\mathrm{KG}}$–C$_{\mathrm{KG}}$ space. By computing SR$_{\mathrm{KG}}$, D$_{\mathrm{types}}$, $H(C)$, $H(R)$, and C$_{\mathrm{KG}}$ at ingestion time and classifying queries into the four tiers, practitioners can apply this study’s findings as a decision framework. In practice, graph data scientists should consider the following recommendations:

\begin{itemize}
  \item \textbf{Document‐Oriented Engines:} they well suit datasets where attribute information dominates and edges are sparse. Document‐oriented engines exploit nested document structures and efficient field indexing, making them ideal for single‐attribute filters and small‐scale lookups. A KG of this kind is a representation where nodes carry most of the information through rich and detailed attributes, while the relationships between them are sparse and semantically weak. The structure is therefore loosely connected but highly descriptive.
  \item \textbf{Multi‐Model Solutions:} they well suit graphs with SR${\mathrm{KG}}$ in the 5–10 range and C${\mathrm{KG}}\approx0.8$, reflecting mixed connectivity patterns where sparse, information‐dense regions coexist with dense clusters. Their hybrid architectures seamlessly blend document retrieval with targeted graph navigation, ensuring efficient one–two hop traversals and attribute‐join operations without full graph‐native overhead. KGs of this type are best described as structurally hybrid, combining regions of high semantic density with areas dominated by attribute-rich but weakly connected nodes.
  \item \textbf{Graph‐Native Platforms:} they should be used when SR$_{\mathrm{KG}}>8$ or C$_{\mathrm{KG}}>0.8$, which indicates richly semantic interconnected subgraphs requiring deep, multi‐hop traversals. Graph‐native platforms provide adjacency‐optimized storage and native traversal planning, minimizing random I/O and leveraging in‐memory path exploration for complex inference and recursive query patterns. A KG of this kind is homogeneously semantic and densely connected, where most regions are rich in relationships and strongly interlinked.
\end{itemize}

\section{Conclusions} \label{sec:conclusions}

KGs management demand more than raw throughput, it requires careful alignment between storage schemas, graph semantics, and connectivity patterns. The semantic richness and connectivity density metrics we introduced reveal meaningful differences in how NoSQL paradigms handle the diversity and sparsity inherent in KGs. Through our empirical evaluation of MongoDB, Neo4j and ArangoDB on the FAERS adverse event graph, we discovered clear trade‐offs that provide meaningful insights for practitioners. 

By establishing a concrete bridge between metric‐driven analysis and empirical performance observations, this work transforms the traditionally ad‐hoc process of DBMS selection into a systematic, evidence‐based methodology. Providing practitioners with concrete guidelines for schema selection that are directly tailored to their KG’s specific semantic characteristics and query requirements, moving beyond intuition‐based decisions toward data‐driven infrastructure choices.

Looking toward the future, this research establishes the foundation for automated, generalized frameworks that can dynamically adapt storage schemas to any KG dataset. Future works aim to develop intelligent systems that automatically select the optimal paradigm through rigorous classification of KG characteristics. This evolution toward intelligent, self‐adapting database selection represents a natural next step in making KG technologies more accessible and efficient for practitioners across different domains.

\begin{acks}
Rosario Napoli is a PhD student enrolled in the National PhD in Artificial Intelligence, XL cycle, course on Health and
life sciences. This work has been partially funded by 
the Horizon Europe ``Open source deep learning platform dedicated to Embedded hardware and Europe'' project (Grant Agreement, project 101112268 - NEUROKIT2E),
the “SEcurity and RIghts in the CyberSpace (SERICS)” partnership (PE00000014), under the MUR National Recovery and Resilience Plan funded by the European Union – NextGenerationEU. In particular, it has been supported within the SERICS partnership through the projects FF4ALL (CUP D43C22003050001) and SOP (CUP H73C22000890001), the Italian Ministry of University and Research (MUR) ``Research projects of National Interest (PRIN-PNRR)'' through the project “Cloud Continuum aimed at On-Demand Services in Smart Sustainable Environments” (CUP: J53D23015080001- IC: P2022YNBHP) and the Italian Ministry of University and Research (MUR) ``Research projects of National Interest (PRIN)'' call through the project ``Tele-Rehabilitation as a Service (TRaaS)'' (CUP J53D23007060006). 
\end{acks}

\bibliographystyle{ACM-Reference-Format}

\bibliography{bibliography}


\end{document}